\documentclass[%
 reprint, 
 amsmath,amssymb,
 aps,
]{revtex4-2}

\usepackage{graphicx}
\usepackage{dcolumn}
\usepackage{bm}
\usepackage{xcolor}

\def\araa{ARA\&A}
\def\apj{ApJ}
\def\apjl{ApJ}
\def\aap{A\&A}
\def\mnras{MNRAS}
\def\na{New A}
\def\prl{Phys.~Rev.~Lett.}
\def\pasj{PASJ}
\def\sovast{Soviet~Ast.}
\def\nat{Nature}

\begin{document}

\preprint{APS/123-QED}

\title{Evidence of a lepto-hadronic two-zone emission in flare states}

\author{E. Aguilar-Ruiz$^1$,  N. Fraija$^1$  and A. Galv\'an-G\'amez$^1$}

\address{$^1$ Instituto de Astronom\'ia, Universidad Nacional Aut\'onoma de M\'exico,
Ciudad de M\'exico, M\'exico,\\
E-mail: eaguilar@astro.unam.mx, nifraija@astro.unam.mx, 
agalvan@astro.unam.mx\\}

\date{\today}

\begin{abstract}
The BL Lac Markarian 501 exhibited two flaring activities in the very-high-energy (VHE) band in May 2009.  The lack of correlation between X-rays and TeV gamma-rays without increasing in other bands suggested that more than one emission zone could be involved. Moreover, fast variability in the flaring state was observed, indicating that the emission zones responsible must have small sizes. We use a lepto-hadronic model with two-zone emission to explain the spectral energy distribution during quiescent and these flaring states.  In the proposed scenario, the photopion processes explain the VHE flaring activities successfully, and variability constraints place the activity in a zone located near the jet's base or named inner blob, while synchrotron self-Compton emission describing the X-ray signature during that flaring state occurs in the zone situated far the central engine or named outer blob.
\end{abstract}

\maketitle

\section{Introduction}\label{aba:sec1}
Blazars are active galactic nuclei (AGN) that host relativistic jets pointing very close to our line of sight \citep{1979ApJ...232...34B}. They are classified in BL Lac objects, which are a sub-set of Flat Spectrum Radio Quasars (FSRQ) \citep{1996MNRAS.281..425M, 2009MNRAS.396L.105G, 2009ApJ...700..597A}. Another subset that might be added to these sources is that with observed neutrinos \cite{IceCube2018Sci...361..147I,IceCube2018Sci...361.1378I}.   Due to the orientation of one of the two jets towards our line of sight, the emission is highly beamed and Doppler boosted, making them bright and variable in all wavebands from radio to $\gamma$-rays \citep{1995ARA_A..33..163W, 1997ARA_A..35..445U}.  These objects show extreme variations in different bands across the electromagnetic spectrum during the quiescent and flaring states, although flaring episodes are still under debate.

Blazars' spectral energy distribution (SED) shows two maxima or peak values in two distinct frequency positions \citep{1995ApJ...444..567P,2002A&A...386..833G, 1996ApJ...463..444S}. The first peak is generally well-fitted, assuming synchrotron emission, and is used further to divide blazars into low, intermediate, and high-peaked synchrotron sources (LSP, ISP, and HSP, respectively). LSP objects are those with the first peak found at frequencies log\,$\nu_{\rm peak}<$\,14, for the ISP sources, the first peak is observed in the range 14\,$<$\,log\,$\nu_{\rm peak}<$\,15 and finally, the HSP ones show their first synchrotron peak at log\,$\nu_{\rm peak}>\,$15. Applying these criteria, the first peak of the broadband SED in the FSRQ objects is observed at the infrared bands; therefore, they are classified as LSP blazars. For the BL Lac class,  the first peak is observed at frequencies that go from the infrared to the hard X-ray bands; i.e., they can be classified as  LSP, ISP, or HSP sources \citep{Abdo_2010}. 

Moreover, BL Lacs that are characterized by having a synchrotron peak located at log\,$\nu_{\rm peak}>\,$17, are commonly named extreme synchrotron peaked (EHSP). In addition, there is a new class of BL Lacs having its high-energy bump located at energy $E_{\rm peak}>\,$1 TeV; they refer as \textit{extreme-TeV} BL Lacs or hard-TeV BL Lacs (TBL) \citep{Tavecchio_2011, Costamante2018}. There is evidence that such extreme behaviors are temporal states and that these conditions are not necessarily simultaneous. For example, Markarian 501, 1ES 1727+502, and 1ES 1741+196 have exhibited these behaviors. \citep{2018A&A...620A.181A,Ahanorian_1999A&A...349...11A, Archambault_2015ApJ...808..110A,Ahnen_2017MNRAS.468.1534A}.

With a low redshift, Markarian 501 (Mrk 501) is one of the closest BL Lac-type blazars to the earth \citep{1996ApJ...456L..83Q}. Due to its distance and low attenuation of gamma rays with cosmic background radiation, Mrk 501 is a crucial astrophysical object to study the high-energy emission process in blazars. Mrk 501 is a very active blazar experiencing many flaring activities since its discovery \citep[see, e.g.;][]{1997ApJ...487L.143C, 1998ApJ...492L..17P, 1999ApJ...514..138K,  1999ApJ...522..846X, 1999ApJ...518..693Q, 2000ApJ...536..742P, 2003ApJ...597..851K, 2004A&A...422..103M, 2006ApJ...646...61G, 2007ApJ...669..862A, 2008NewA...13..375G, Abdo_2011ApJ...727..129A, 2015A&A...573A..50A, 2015ApJ...812...65F, 2016A&A...594A..76A, 2016A&A...593A..91A, 2018A&A...620A.181A, Ahnen2017A&A...603A..31A, MAGIC_2020A&A...637A..86M}.

For instance, in May 2009, on the 1st and 22nd, it presented two flare episodes. These flares were studied by \citep{Ahnen2017A&A...603A..31A}, who emphasized the difficulty of modeling the 22nd of May flare with a one-zone synchrotron self-Compton (SSC) scenario and suggested a better description invoking two independent zones under the SSC model.  They also concluded that the very-high-energy (VHE; $\geq 100\,{\rm GeV}$) gamma-rays could come from a variable component that could contribute to the produced emission by the SSC emitting zone, which must be responsible for the dominant X-ray emission. A similar model was proposed by \citep{Lei_2018PASJ...70...45L}, who pointed out that these two flaring stages could be explained by invoking the interaction of two emission zones (a gamma-ray and a radio-emitting zone). 

In this work, we describe these flares using the recent two-zone lepto-hadronic model proposed by Aguilar-Ruiz et al. \citep{Aguilar_Ruiz_2022arXiv220300880A} (hereafter AR2022) introduced to explain the broadband emission of hard TeV BL Lacs.  In the AR2022 model, the maxima of the SED's bumps are produced in different dissipation regions. The low-energy bump is governed by the synchrotron emission of accelerated electrons confined in the outer blob region; meanwhile,  the high-energy bump is produced by the decay of neutral pions resulting from the photopion process.  We explore the possibility that the AR2022 model could explain the flaring activities exhibited in Mrk 501. The structure of this manuscript is organized as follows. Section 2 introduces the theoretical model that describes the VHE gamma rays. In Section 3, we consider the multi-wavelength observations of Mrk 501 around the flare activity in 2009, and finally, we discuss and summarize in Section 4.
%
%
\section{Theoretical model}\label{sec:Model} 
The model proposed by AR2022 to describe the entire spectral energy distribution of six TBL during their quiescent state involves two-emission zones. The authors required two dissipation regions named inner and outer blob to relax the parameter demanded by the one-zone SSC model. Meanwhile, the outer blob lies far from the central engine; the inner blob is near the jet's base. Additionally, the author used the possible formation of a pair plasma that emerges and is launched above the accretion disc. This plasma pair generates a narrow shape spectrum with characteristic energy centered $\varepsilon_{\rm pl} = 511 \, \rm keV$.  While this annihilation line has not been observed in blazars, it has recently been claimed that it was observed during a significant flare in the microquasar V404 Cygni \citep[see,][]{Siegert_2016Natur.531..341S}. Observation of this emission may not be exclusive to microquasars; if there exists a universal relation for accreting black holes at all scale masses, as many authors have suggested \citep[e.g., see][]{Falcke_2004A&A...414..895F,Ploktin_2012MNRAS.419..267P, Markoff_2010LNP...794..143M}, this line could also be expected during the flaring activities in blazars.
Both relativistic electrons and protons in the inner blob scatter off photons produced by the pair plasma. While protons interact via photohadronic processes (i.e., photopion and photopair),  electrons do via the Compton scattering mechanism. The photopion process produces neutral pion, which decays into gamma-rays with energies around $\sim 1 \rm \, TeV$. Meanwhile, both primary and secondary electrons cool down via synchrotron radiation with a signature at radio-to-optical and MeV bands, respectively; electrons in the inner blob could also have a signature at MeV produced by external Compton scattering.   The fluxes produced in the MeV band are not dominant because these contributions are strongly attenuated below the GeV band by the pair-plasma radiation (see figure 2(b) in AR2022). Furthermore, the rest of the emission is described by a SSC model in the outer blob produced by relativistic electrons; proton emission is irrelevant in this blob. 

It is essential to mention that the main results of AR2022 for TBLs are i) the model avoids the Klein-Nishina flux suppression in the outer blob against the one-zone SSC model present due to the very high electron's Lorentz factor, ii) the equipartition value is close to the unity $U_B/U_e \gtrsim 0.1$ in the outer blob against the very low value demanded by the one-zone SSC model $U_B/U_e \ll 1$, and iii) the minimum electron Lorentz factor is around $\sim 50$ in contrast to the extreme value $\gtrsim 10^3$ required by the one-zone SSC model.

We use the same reference frames as AR2022 for the observed, the pair plasma, and the blob (inner or outer) frame. We employ Latin Capital Letters with the superscript ``{\rm ob}" for observed quantities, while the AGN frame will be without the superscript. Lowercase letters with unprimed, prime, or two-prime are used in Greek for the plasma, inner, and outer blobs. For instance, the observed energy is $E^{\rm ob}$, and the energy measured in the comoving frame of the AGN, the pair-plasma, the inner blob, and the outer blob are $E$, $\varepsilon$, $\varepsilon^\prime$, $\varepsilon^{\prime\prime}$, respectively.  Furthermore, we consider an on-axis case for relativistic blobs with a viewing angle of $\theta_{\rm obs} \lesssim 1/\Gamma$ and the Doppler factor defined by $\mathcal{D} = \left[\Gamma(1-\beta\cos{\theta_{\rm obs}})\right]^{-1} $, with $\Gamma$ the Lorentz factor in the blob. 
In the following, we describe the AR2022 model applied to the quiescent state and during the flaring state in more detail.

\subsection{Quiescent state}\label{sec:MODEL_quiescent}
We use the AR2022 model to explain the broadband emission in both the quiescent and the flaring states of Mrk 501 in May 2009. Meanwhile, the quiescent state is well explained using similar assumptions as did in AR2022; the flaring activities must be interpreted into a more complex scenario which will be discussed below in the section \ref{sec_flare_state_model}.

We summarize the main features of the AR2022 model taken into account in this work:  
\begin{itemize}
    \item [1.] The outer blob: i) The electron population is the dominant component compared with the proton one, $n_e \gg n_p$; then, only leptonic processes are considered,   ii) the blob moves with relativistic speed such that its Lorentz and Doppler boost factors are $\Gamma_o = 5$ and $\mathcal{D}_o \simeq 10$, respectively. iii) Finally, the blob's size is constrained using the variability timescales $R''_o \simeq \mathcal{D}_o t_{\rm var} c$ assuming $t_{\rm var} \sim {\rm one\,day}$ and $\mathcal{D}_o\sim 10$, and the blob's location from the super massive black hole (SMBH) is $r_o = \Gamma_o R''_o \sim 10^{17} \, \rm cm$, which represents about $\sim 10^3 R_g$ with $R_g= GM_\bullet/c^2 \sim 10^{14} \, \rm cm$ the Schwarzschild radius for a SMBH mass of $M_{\bullet} = 10^{9} M_{\odot}$. Note that this distance is very similar to the suggested acceleration and collimation zone \citep[e.g][]{Marscher_2008,Walker_2018ApJ...855..128W}, therefore, $\Gamma_o$ could have a value close to the terminal Lorentz factor of the jet. Additionally, It is worth noting that \cite{Daly_2019ApJ...886...37D} has argued that such radio galaxies pointed at us typically have a highly spinning BH, allowing to use Blandford \& Znajek's work \cite{Blandford_1977MNRAS.179..433B}.
    \item [2.] The inner blob: i) It is assumed to have one electron per proton, $n_e=n_p$ and the same spectral index for electrons and protons $\alpha_e = \alpha_p$.With these conditions, the electron and proton luminosities are given by $L_e \simeq L_p$ \citep{Kardashev_1962SvA.....6..317K}. ii) The location must be closer to the central engine than the outer blob in order to be influenced by the radiation field of the pair-plasma, around dozens of $R_g$, and iii) the blob's Lorentz factor determines the peak of VHE band emission, which is favorable to mildly relativistic speed. Generally, in our scenario, the Lorentz factor in the inner region is less or equal to the outer one ($\Gamma_i \lesssim \Gamma_o$), which is feasible if two regions are inside the acceleration and collimation zone.
    \item[3.] The pair plasma: i) It moves with a mildly relativistic velocity at the photosphere $\beta_{\rm pl} = 0.5$ ($\Gamma_{\rm pl}= 1.15$), ii) we take the lower disc's luminosity above $\rm 511 \, {\rm keV}$ to guarantee the formation of the outflow, $L_{\rm keV}\approx 3 \times 10^{-3} \, L_{\rm E}$, where $L_E \approx 1.26 \times 10^{47} \, {\rm erg\, s^{-1}}$ is the Eddington's luminosity corresponding to the SMBH's mass $M_{\bullet} = 10^9 \, M_{\odot}$ such that harbours Mrk 501 \citep{Wagner_2008MNRAS.385..119W}.
    \item[4.] The outer blob is far away from the pair plasma. Therefore, its radiation field is irrelevant for outer blob's processes.
    \item[5.] The magnetic field is estimated, assuming the magnetic energy is conserved along with the jet. Therefore, the value in the inner and in the outer blob could be related as 
    \begin{equation}\label{eq_magnetic_conservation}
        B_i = (r_o/r_i) \, B_o \, .
    \end{equation}
    \item [6.] The electron(proton) energy break of each distribution could be estimated equaling the synchrotron and the adiabatic loss timescales. Therefore, it reads as 
    \begin{equation}\label{eq_gamma_br}
        \gamma_{e(p), \rm br} = \frac{3 \pi m_{e(p)}^3 c^2}{\sigma_T m_e^2 B^2 R} \,. 
    \end{equation}
    The maximum energy is estimated from the competition between the acceleration and loss processes. When the adiabatic losses are dominant we have 
    \begin{equation}\label{eq_gamma_max}
        \gamma_{e(p), \rm max} = \alpha_{acc}  \frac{eB R}{m_{e,p} c^2} \, ,
    \end{equation}
    %
    where $\alpha_{\rm acc}$ is the acceleration efficiency, in this work we take a value of $\sim 0.1$ \citep{Caprioli_2012}.

    \item [6.] AR2022 considered the pair-plasma's photons are redshifted when they are observed in the inner blob.  
    Therefore, the energy and energy density measured in the inner blob are 
\begin{equation}\label{eq_plasma_energy_redshift}
   \varepsilon'_{\rm pl} \simeq \varepsilon_{\rm pl}/(2\Gamma_{\rm rel}) \, , \quad {\rm and} \quad
    u^{\prime}_{\rm pl}  \simeq u_{\rm pl}/(2\Gamma_{\rm rel})^2 \, ,
\end{equation}
respectively. 
where $\Gamma_{\rm rel}$ is the relative Lorentz factor between the pair plasma and the inner blob, which can be expressed as
\begin{equation}\label{eq_GammaRel}
\Gamma_{\rm rel} = \Gamma_{\rm i} \Gamma_{\rm pl} \left( 1 - \beta_{\rm i} \beta_{\rm pl} \right)\,.
\end{equation}
The terms $\beta_i$ and $\Gamma_i$ are inner blob's velocity and its respective Lorentz factor, respectively. For practice, the Doppler boost and Lorentz factors are related as $\mathcal{D}_i=2\Gamma_i$.
\item[7.] The gamma-ray peak from neutral pion decay product is estimated by relativistic kinematics.  Then, it can be written as 
\begin{equation}\label{eq_pion_th_gamma}
    E_\gamma^{\rm ob} \simeq 1.6 \times 10^{16} \, {\rm eV}^2 \, \mathcal{D}_i \, \varepsilon_{\rm pl}'^{-1}
\end{equation}
\item[8.] Here, in contrast with AR2022, the broadline (BLR) and dusty torus (DT) regions are not considered. It is worth noting that although many authors have suggested different BLR luminosities for Mrk 501, e.g., $L_{\rm BLR}\approx 1.6 \times 10^{42} \rm \, erg \, s^{-1}$ \citep{Sbarrato_2012MNRAS.421.1764S} or $L_{\rm BLR}\approx 5.2 \times 10^{40} \rm \, erg \, s^{-1}$ \citep{Stocke_2011ApJ...732..113S},  all of them are lower than those ones assumed by AR2022 for the TBLs considered therein , i.e, $L_{\rm BLR} \approx 2\times 10^{\rm 43} (2\times 10^{\rm 42}) \rm \, erg \, s^{-1} \, $ for $M_{\bullet}=10^9 (10^8) M_{\odot} $.
\item[9.] The host galaxy produces the optical-UV bump emission, which is not included in our model.
\item[10.] The model proposed in AR2022 does not explore the emission of secondary electrons in detail because when pair-plasma photons are redshifted into the inner blob frame, the emission of secondary pairs peaks near the MeV band. At that energy, the identical pair-plasma photons strongly attenuate the flux. Nevertheless, as we discussed before, the observed flux must be partially absorbed below MeV energies. In the following, we discuss the emission of secondary pairs.
\end{itemize}

\subsubsection{Secondary pairs}
The flux of synchrotron secondary pairs could be observed, especially during an intense flaring state where the VHE gamma-ray flux increases. In an environment such as the inner blob, where electrons are efficiently cooled down via the synchrotron mechanism, the produced fluxes via photopion and photopair could be related to proton luminosity as $L_{\gamma,p\pi} \approx (1/8) f_{p\pi} L_p$ and $L_{\gamma, pe} \approx f_{pe} L_{p}$, respectively (see \citep{Fraija_2020MNRAS.497.5318F, Petropoulou_2015MNRAS.447...36P}). It is worth noting that a flux ratio of both processes become $\frac{L_{\rm \gamma,p\pi}}{L_{\rm \gamma,pe}}\sim \frac{\sigma_{\rm p\pi}^{\rm pk}}{8/\sigma_{\rm pe}^{\rm pk}} \sim 10^2$, where $\sigma_{\rm p\pi}^{pk}$ and $\sigma_{\rm pe}^{\rm pk}$ are the cross-sections of photopion and photopair processes, respectively.    Therefore in our model, we consider only the emission of photopion pairs.    An estimation of the energy peak of photopion pairs could be done considering that the average energy transferred from the proton to the electron is $\varepsilon_e/\varepsilon_p \approx 0.05$.  From kinematics, the minimum electron's Lorentz factor is hence estimated by the photopion proton threshold
\begin{equation}
    {\gamma'}_{e,p\pi}^{\rm th} \gtrsim 1.6 \times 10^{10} \, \left( \frac{ \varepsilon_{\rm pl}'} {\rm eV} \right) ^{-1} \, .
\end{equation}
Considering typical magnetic field in the inner blob we obtain the peak of synchrotron emission
\begin{equation}\label{eq_Esyn_photopion}
    E_{\rm syn,p\pi} \gtrsim 15 \, {\rm keV \, MeV^2} \, \varepsilon_{\rm pl}'^{-2} \, 
    \left(\frac{B_i'}{100 \rm G}\right) \, \left(\frac{\mathcal{D}_i}{5}\right) \, .
\end{equation}
It is essential to mention that in the photopion scenario, the gamma-ray and  secondary-pair fluxes are related as $L_{\rm syn} \approx (1/4) L_{\gamma}$ \citep{Ahlers_2017PTEP.2017lA105A}. For instance, the VHE emission in TBLs is $L_{\gamma}^{\rm VHE} \sim 10^{45} \, \rm erg \, s^{-1}$, so that this would imply a synchrotron luminosity of $L_{\rm syn}^{p\pi}
\sim 2.5 \times 10^{44} \, \rm erg \, s^{-1}$.

\subsection{Flaring states}\label{sec_flare_state_model}
We require a more complex description to interpret flaring activities. A flaring state is defined as the luminosity in one or more electromagnetic spectral bands increasing during a short period. The photon luminosities usually depend on timescales of radiative processes, the bulk Lorentz factor of the jet, the electron (proton) luminosities, etc. For instance, the observed luminosity resulting from neutral pion decay produced in the photopion process is
\begin{equation}
    L_{\gamma, p\pi}^{\rm ob} \propto {t'}_{p\pi}^{-1} \, {L'}_{p} \, \Gamma_i^4 \, ,
\end{equation}
where ${t'}_{p\pi}$ is the photopion loss timescale and $L_p'$ is the proton luminosity in the comoving frame.   Similarly, the synchrotron luminosity from secondary pairs produced from  photopion and photopair process is
\begin{equation}
    L_{\gamma,pe (p\pi)}^{\rm ob} \propto  {t'}_{e, \rm syn}^{-1} {t'}_{pe (p\pi)}^{-1} \, L'_{p} \, \Gamma_i^4 \, ,
\end{equation}
where ${t'}_{e, \rm syn}$ and ${t'}_{pe}$ are the synchrotron loss timescale for electrons and the photopair loss timescale, respectively. The proton synchrotron must be considered when the magnetic field is strong enough, as expected in the inner blob. In this case, the luminosity becomes
\begin{equation}
    L_{p,\rm syn}^{\rm ob} \propto {t'}_{p, \rm syn}^{-1} \, L'_{p} \, \Gamma_i^4 \, ,
\end{equation}
where ${t'}_{p,\rm syn}$ is the proton synchrotron loss timescale. Primary electrons cool down mainly by synchrotron and Compton scattering mechanisms which are given by
\begin{equation}
    L_{e,\rm syn}^{\rm ob} \propto {t'}_{e, \rm syn}^{-1} \, L'_{e} \, \Gamma_i^4 \, ,
\end{equation}
and
\begin{equation}\label{n1}
    L_{e,\rm IC}^{\rm ob} \propto {t'}_{e,\rm C}^{-1} \, L'_{e} \, \Gamma_i^4 \, ,
\end{equation}
respectively, ${t'}_{e,\rm C}$ is Compton loss timescale for electrons.

Note that all observed luminosities depend directly on three quantities: i) the losses timescales, ii) the electron/proton luminosity, and iii) the blob's Lorentz factor. Therefore,  the increase of one of them guarantees an increase in the observed luminosity, which could be associated with a flaring state.

\subsubsection{Enhance of the processes efficiency}

The evolution of the cooling timescale gives us information about the efficiency of the process. In our model, the cooling timescales $t_{p\pi}^{-1}$, $t_{pe}^{-1}$ and $t_{e,\rm EC}^{-1}$  are directly proportional to the external seed photons, meanwhile timescale $t_{e(p), \rm syn}^{-1}$ is proportional to the strength of the magnetic field.

In our model, the external seed photons are provided by the pair plasma, and therefore, the energy density is
\begin{equation}\label{eq_pair_plasma_density}
    u_{\rm pl}' \simeq \frac{L_{\rm keV}  }{\Omega_{\rm pl} R_{\rm ph}^2 \beta_{\rm pl } c\, 4 \, \Gamma_{\rm rel}^2}   \,  \,,
\end{equation}
where $\Omega_{\rm pl}\sim 0.2 \pi$ is the solid angle covered by the pair plasma, and $R_{\rm ph}\sim R_g$ is the radius at photosphere.
Following the treatment derived in \citep{Kelner&Aharonian_2008} and using equation (\ref{eq_pair_plasma_density}), the photopion timescale in delta-approximation is written as
\begin{align}\label{eq_photopion_timeloss}
    {t'}_{p\pi}^{-1} (\gamma'_p)
    &= \frac{ L_{\rm keV}}{2 \Gamma_{\rm rel} \Omega_{\rm pl} R_{\rm ph}^2 \beta_{\rm pl } \varepsilon_{\rm pl} } 
    \int_0^1 dx \, x \, \Phi_{\rm all} \left(\frac{4\gamma_p' \varepsilon_{\rm pl}'}{m_p c^2}, x \right)\,.
\end{align}
On the other hand, following \citep{Chodorowski_1992} the photopair timescale can be written as
\begin{equation}\label{eq_photopair_timesloss}
{t'}_{\rm BH}^{-1}(\gamma_p') =  
\frac{3 \sigma_{\rm T} \alpha_f m_e^3 c^4}{32 \pi m_p }  \frac{ L_{\rm keV}}{\Gamma_{\rm rel}\Omega_{\rm pl} R_{\rm ph}^2 \beta_{\rm pl } \varepsilon_{\rm pl}^3 }
\frac{1}{\gamma_p'^3}  \varphi \left( \frac{2 \gamma_p' \varepsilon_{\rm pl}'}{m_e c^2} \right) \, ,
\end{equation}
where  $\varphi$ is a parametrized function.  
Similarly, the Compton scattering loss timescale in delta-approximation is
\begin{equation}\label{eq_IC_timesloss}
{t'}_{\rm IC}^{-1} (\gamma_e') 
\simeq\frac{3 \sigma_{\rm T} }{8 m_e c} \frac{ L_{\rm keV}}{\Gamma_{\rm rel}\Omega_{\rm pl} R_{\rm ph}^2 \beta_{\rm pl } \varepsilon_{\rm pl}^2 } \frac{1}{\gamma_e'^3}\int dE_1 E_1 F_c(q,\Gamma_e) \, ,
\end{equation}
where $q$ and $\Gamma_e$ are given by \citep{BLUMENTHAL&GLOUD_1970}.

We notice from the above equations that when $L_{\rm keV}$ increases, the efficiency of photopion, photopair, and EC processes enhance but do not change any spectral signatures, which can also be observed when flux increases.

Belodoborov \citep{Beloborodov1999MNRAS.305..181B} proposed the formation of a pair plasma, which could emerge above the accretion disc if the luminosity $L_{\rm keV}$ increases enough to produce a thick optical environment to create annihilate pairs.  These processes produce an $e^\pm$ outflow that moves with a mildly relativistic velocity of $\beta_{\rm pl} \approx 0.3 - 0.7$. For BL Lacs and during a quiescent state, the accretion disc luminosity cannot be higher than $L_d \approx 5 \times 10^{-3} L_{E}$ \citep{Ghisellini&etal_2011}. This luminosity condition may not be hold during flaring states.

Moreover, we note that efficiencies are also a function of $\Gamma_{\rm rel}$, but as Eq. (\ref{eq_GammaRel}) shows, could be assumed only dependent of $\Gamma_i$ because $\beta_{\rm pl}$ cannot take a wide range of values. The case where $\Gamma_i$ increase is discussed in a next subsection.

\subsubsection{Variation in the particle distribution}

We assume that accelerated protons reach a steady state during flaring events, and also, the proton distribution is isotropic and homogeneous \footnote{This refers to a comoving frame although is not a necessary condition \cite{Lind&Blandford_1985ApJ...295..358L}}. Then, proton distribution follows a single power-law (PL) function given by  
\begin{equation}\label{eq_pDistribution}
{N_p'}({\varepsilon_p'}) = {K_p'} {\varepsilon'_p}^{-\alpha_{p}} \, 
\qquad \varepsilon'_{p, \rm min} \leq \varepsilon_p' \leq \varepsilon'_{p, \rm max}\, ,
\end{equation}
where the term $K_p'$ is the normalization constant, $\alpha_p$ is the proton spectral index and ${\varepsilon}'_{p, \rm min}$ and ${\varepsilon}'_{p, \rm max}$ are the minimum and maximum energy, respectively. 
Similarly, the electron distribution is described by a broken PL as
{\small
\begin{equation}\label{eq_eDistritbution}
N_e'(\gamma_e') = K_e'
\begin{cases}
{\gamma'_e}^{-\alpha_{e,1}}, 
\qquad \qquad \qquad \quad \;\; {\gamma'_{e, \rm min}} \leq {\gamma'_e} \leq {\gamma'_{e, \rm br}} 
\\
{\gamma'_{e, \rm br}}^{\alpha_{e,2}-\alpha_{e,1}} {\gamma'_e}^{-\alpha_{e,2}}, 
\qquad {\gamma'_{e, \rm br}} \leq {\gamma'_e} \leq {\gamma'_{e, \rm max}}
\end{cases}
\end{equation}
}
where ${\gamma'}_{e, \rm min}$, ${\gamma'}_{e, \rm br}$ , ${\gamma'}_{e, \rm max}$ are the minimum, break and maximum Lorentz factors of ultrarelativisc electrons, respectively.

Variations in the electron/proton distribution are suggested to originate in flaring states. For instance, \citep{Mankuzhiyil_2012ApJ...753..154M} modeled different states of Mrk 501, assuming they are a consequence of variations of the electron distribution inside the emission's zone. They pointed out that an intense and softer distribution produces a quiescent state, while a distribution with a harder spectral shape gives rise to a flaring state. Furthermore, increased injection of particles inside the emission zone could also trigger flare episodes \citep[e.g., see][]{Bottcher_2019ApJ...887..133B}. 

We consider two approaches for which  the particle distribution could vary in a flaring episode: i) when the electron/proton distribution gets hardened, but the total particle number is conserved, $N_{e,p}^{\rm flare} = N_{e,p}^{\rm quiescent}$, i.e., there is no injection of new particles but an acceleration process, we refer it as \textit{re-acceleration case},  ii) when the distribution gets hardened or remains equal, but the total particle number increases $N_{e,p}^{\rm flare} > N_{e,p}^{\rm quiescent}$, we refer it as \textit{injection case}.   It is worth mentioning that the mechanism of how the particle distribution could get hardened or injected is unclear, and it is out of the scope of this work.  Here, we use distinct parameters for describing different states.

\paragraph{Particle re-acceleration.}

Assuming particle number is conserved, the normalization constant varies concerning the previous state, only as a function of spectral indexes, by a factor of {\small $K_{e,p}^{\rm flare} \simeq (\alpha_{e,p}^{\rm flare}-1)/(\alpha_{e,p}^{\rm quiescent}-1) K_{e,p}^{\rm quiescent}$}. We note that the maximum and break energies do not have relevance. Therefore, the electron/proton luminosity during the flare becomes
\begin{equation}
    L_{e,p}^{\rm flare} \simeq 4\pi R^2 \, c \,  \Gamma^4 u_{e,p} > L_{e,p}^{\rm quiescent} \,.
\end{equation}
The previous inequality (eq. 20) is valid when $\alpha_{e,p}^{\rm flare} < \alpha_{e,p}^{\rm quiescent}$, and also when $\varepsilon_{p, \rm max}$ and $\gamma_{e, \rm br}$ do not decrease during the flaring activity. 
\paragraph{New particle injection.}
New electrons and protons could be injected with a different intensity and spectral shape than the existing ones when the blobs are quiescent. The difference with the re-acceleration case is that this does not conserve the total particle number.  This implies that the luminosity during a flaring state is larger than in a quiescent state, $L_{\rm e,p}^{\rm new} > L_{\rm \rm e,p}^{\rm old}$, as observed.   In the AR2022 model, particles could be injected inside one blob at once or inside both blobs simultaneously. Moreover, we assume particle distributions reach a steady state during the flare duration. 

Another important assumption is that the neutrality condition between injected protons and electrons inside the inner blob is satisfied \citep{2013ApJ...768...54B, 2017APh....89...14F}. It is worth noting that this condition is not needed for the outer blob. Finally, particle distribution changes (e.g., a spectral index) must reflect variations in the SED,  as observed in a flaring activity.

\subsubsection{Acceleration of the emission zone}\label{sec:MODEL_acceleration_case}
 As a blob accelerates, the kinetic luminosity increases ($L_{\rm k} = 4\pi R^2 \, c \, \Gamma_b^4 \, u_{\rm k}$ with {\rm k=e, p, rad and B}) if no other parameter changes.   

Another consequence is the shifting of observed energy peaks to higher values ($E^{\rm ob} = \mathcal{D} \varepsilon$).  Moreover, using Eqs. (\ref{eq_pion_th_gamma}) and (\ref{eq_GammaRel}) the threshold gamma-ray energy must be greater than
\begin{equation}
    E_\gamma^{\rm ob} \gtrsim 64 \, {\rm GeV} \, \mathcal{D}_i \, \Gamma_{\rm i} \Gamma_{\rm pl} \left( 1 - \beta_{\rm i} \beta_{\rm pl} \right) \approx 90 \, {\rm GeV} \, \Gamma_i^2 \,.,
\end{equation}
where we have used $\beta_{\rm pl}=0.3\, (\Gamma_{\rm pl}=1.05)$ and $\mathcal{D}_i \simeq 2\Gamma_i$,  for $\beta_i\approx 1$.
Therefore, the observed spectrum shifts to higher energies as $\Gamma_i$ increase. 
An important implication is that the increase of the $\Gamma_i$-value must produce an increase of the photopion, photopiar and EC timescales, although the efficiencies decrease (see Eqs.  \ref{eq_photopion_timeloss}, \ref{eq_photopair_timesloss} and \ref{eq_IC_timesloss}). 

Therefore, a flare activity under the AR2022 scenario could be interpreted by: i) an increase on the pair-plasma luminosity, ii) an increase of proton/electron luminosity due to an injection or a hardening of its distribution, and iii) an acceleration of the blob.

%
%
%

\section{Markarian 501: Flaring events in  2009 May}\label{sec:Results}

We apply our model to Mrk 501 in May 2009 for a quiescent state and two flares presented on May first and 22nd, respectively. First, we model the quiescent state, and after, we interpret the flares activities evoking the model described in Section 2.

\subsection{The quiescent state}

We consider the multifrequency campaign performed from march 15, 2009, to August 1 (4.5 months), excluding the flaring activity on May 22 \citep[For details, see ref.][]{Abdo_2011ApJ...727..129A}. During this campaign, Mrk 501 exhibited low activity at all wavebands. We evoke the same treatment shown in \cite{Aguilar_Ruiz_2022arXiv220300880A} to model the low activity exhibited in Mrk 501. As follows,  we estimate the parameter values required by our model.

\subsubsection{The outer blob}
We assume that the outer blob moves relativistically with a Lorentz factor of $\Gamma_{\rm o} = 5$ where for a jet observation in face-on point viewed, the Doppler factor is $\mathcal{D}_{\rm o} \simeq 10$.  The region size could be restricted using the observed variability timescale, $R\simeq \mathcal{D} t_{\rm var} c$. Multifrequency observations indicate that the variability timescales vary between 5 and 10 days \cite{Abdo_2011ApJ...727..129A}, except at VHEs, in which the authors reported values of one day or even shorter.  \cite{Kataoka_2001ApJ...560..659K} found similar results when performing an X-ray analysis of the previous epochs of Mrk 501 and another TeV Blazars.  Here, we constrain the size of the outer region using a variability timescale of days, which corresponds to the size in the blob frame of 

\begin{equation}
    R''_{\rm o} \lesssim 2.6 \times 10^{16} \, {\rm cm} \, \left( \frac{t_{\rm var}}{1 \, \rm day} \right) \left( \frac{\mathcal{D}_{\rm o}}{10}\right) \, .
\end{equation}

The magnetic field could be estimated using the synchrotron bump's observed peak, around 1 keV. 
 Considering the synchrotron spectral break $E_{\rm syn, br}= eB\gamma_{e, \rm br}^2 \mathcal{D}/(m_e c)$ and Eq. (\ref{eq_gamma_br}), the strength of the magnetic field is
\begin{equation}
    B_{\rm o}'' \sim 0.09 {\, \rm G} \left( \frac{E_{\rm syn}^{\rm ob}}{1 \, \rm keV} \right)^{-1/3} \left( \frac{\mathcal{D}_{\rm o}}{10} \right)^{1/3} \left( \frac{R_{\rm o}''}{1 \times 10^{16}\, \rm cm} \right)^{-2/3} \, .
\end{equation}

The parameters that described the electron distribution can be constrained using the blob's parameters previously estimated. The break Lorentz factor can be approximated using Eq. (\ref{eq_gamma_br}), and then written as

\begin{equation}\label{eq_gammae_br_outer}
    \gamma''_{e, \rm br} \sim 8 \times 10^{4}  \, \left( \frac{B''_o}{0.1 \, \rm G} \right)^{-2} \left( \frac{R_o''}{1 \times 10^{16} \rm \, cm} \right)^{-1} \,.
\end{equation}

The minimum ($\gamma''_{e, \rm min}$) and maximum ($\gamma''_{e, \rm max}$) Lorentz factors as well as the spectral indexes are determined by fitting the broadband SEDs.  The electron normalization constant is determined by the ratio of Compton to synchrotron luminosities, which for Mrk 501 a feasible value is $L_{\rm ic}/L_{\rm syn} \sim 1$, and considering a spectral index of $\alpha_{e,1} = 2.3$, which agrees with observations. Therefore, the electron normalization constant becomes
\begin{align}\label{eq_Ke}
    K_e'' \sim 3 \times 10^4 \, {\rm cm^{-3}} \, \left( \frac{R''_o}{1 \times 10^{16} \, \rm cm} \right)^{-1} \,  \left( \frac{\gamma''_{e, \rm b}}{2 \times 10^{5}} \right)^ {-0.7} \, \,.
\end{align}
The complete set of parameters that describes the outer blob is listed in Table \ref{tab_model_outer}, while the result of SSC flux is plotted in Figure \ref{fig_SED_quiescent}.  We note that only considering the SSC model in the outer blob the VHE observations  is explained by demanding a very low equipartition value. Then, the AR2022 model offers a solution to this issue, invoking the existence of another blob located near the jet's base. In this blob, protons interact via the photopion process with the radiation field produced by a pair plasma that emerges above the accretion disc.

\subsubsection{The inner blob}
The inner blob description is more complex than the outer blob because it involves electrons and protons. {In this work, the inner blob is assumed electrically neutral. We estimate the parameters that describe the inner blob as we did for the outer blob. We assume the size of this blob is of the order of $R_g$. Here, we adopt $R=1 \times 10^{14} \rm cm$ similar to \citep{2011ApJ...736..131A}.  As shown in AR2022, the main emission of this blob corresponds to the highest gamma-rays, which could reach energies of a few TeVs. This flux results from neutral pion decay into two gamma-rays, in which spectrum peaks around $\sim$ 1 TeV.\\

We consider the photons produced in the pair plasma, which reach the inner blob.   As shown in section  \ref{sec:MODEL_acceleration_case}, in this case, the resulting energy peak of gamma rays is independent of $\Gamma_i$ or any other parameter. Therefore, we cannot constrain any parameter, so we take them as free parameters.

Concerning the strength of the magnetic field, we estimate its value using Eq. (\ref{eq_magnetic_conservation}). It can be written as
\begin{equation}\label{eq_magnetic_field_jet}
    B_i' \sim 100 \, {\rm G} \left( \frac{B_o}{0.1 \, {\rm G}} \right) \left( \frac{r_o}{10^{17} \rm \, cm} \right)^{-1} \left( \frac{r_i}{10^{14} \rm \, cm} \right) \,.
\end{equation}
As follows, we estimate the parameters that describe proton and electron distributions.
\paragraph{Proton distribution.}\label{sec_Ep_max}
We consider the minimum proton energy as $\varepsilon_{p, \rm min}'= m_p c^2\simeq 1 \, \rm GeV$, and the maximum energy (Eq. \ref{eq_gamma_max}) as

\begin{equation}
    \varepsilon_{p, \rm max}' \sim 280 \, {\rm PeV \, } \left( \frac{\alpha_{acc}}{0.1} \right) \, \left( \frac{B}{100 \, \rm G}\right) \left( \frac{R'_i}{10^{14} \, \rm cm}\right)\,.
\end{equation}
Therefore, the maximum energy of proton-synchrotron photons is given by 
\begin{equation}
    E_{p, \rm syn}^{\rm ob} \sim 1.3 \, {\rm GeV} \, \left( \frac{\mathcal{D}_i}{3}\right) \left( \frac{B'_i}{100 \, \rm G}\right) \left( \frac{\varepsilon'_{p, \rm max}}{300 \, \rm PeV}\right)^2 \,.
\end{equation}

The observed luminosity at the peak produced by proton synchrotron with $\alpha_p=2$ could be estimated by taking the maximum proton energy as

\begin{multline}
    L_{p, \rm syn}^{\rm ob,pk} \sim 3 \times 10^{43} \, {\rm erg \, s^{-1}} 
    \left( \frac{\mathcal{D}_i}{3} \right)^4 \left( \frac{B'_i}{100 \, \rm G} \right)^2
    \\
    \left(\frac{K'_p}{1 \times 10^{7} \rm \, GeV^{-1} \, cm^{-3}} \right) 
    \left( \frac{\varepsilon'_{p, \rm max}}{300 \, \rm PeV} \right) \, .
\end{multline}

This flux must be attenuated by the photons of the pair-plasma, as pointed out by \citep{Aguilar_Ruiz_2022arXiv220300880A} for TBLs. Moreover, it is worth noting that if we take the lowest value allowed by our model to works, i.e., $\varepsilon'_{p,\rm max} \sim 100 \, \rm TeV$, the observed luminosity would be $L_{p, \rm syn}^{\rm ob,pk} \sim 1 \times 10^{40} \, {\rm erg \, s^{-1}}$.

Therefore, proton-synchrotron contribution may not be taken into consideration at these protons' energies ($\lesssim 100 \, \rm PeV$).

 However, only when $\varepsilon'_{p,\rm max}$ reaches higher values, for instance, $\sim 5 \, \rm EeV$, the synchrotron peak shifts to $E_{p, \rm syn}^{\rm ob} \sim 325 \,\rm GeV$ and the pair-plasma radiation field does not completely suppress the flux at those energies; therefore a significant fraction of proton synchrotron emission must be observed for $\varepsilon'_{p,\rm max} \gtrsim 5 \, \rm EeV$. 

%
\paragraph{Electron distribution.} We consider the minimum and maximum electron Lorentz factors as $\gamma_{e, \rm min}=1$ and $\gamma_{e,\rm max}=10^5$ \citep{2011ApJ...736..131A}, respectively. Note that, against protons, the break Lorentz factor of electrons is located at lower values because electrons cool down quickly due to a strong magnetic field. Therefore, the break Lorentz factor is
\begin{equation}\label{eq_e_break_inner}
    \gamma'_{e, \rm br} \approx 7.5  \, \left( \frac{B'_i}{100 \, \rm G} \right)^{-2} \left( \frac{R_i'}{10^{14} \rm \, cm} \right)^{-1} \, ,
\end{equation}
which produces a synchrotron spectral break  at radio frequencies 
\begin{equation}  
    \nu_{e, \rm syn, pk}^{\rm ob} \approx 6.8 \, {\rm GHz} \, \left( \frac{\mathcal{D}_i}{3} \right) \left( \frac{B_i'}{100 \, \rm G} \right)^{-3} \left( \frac{R_i'}{10^{14} \rm \, cm} \right)^{-2}\,.
\end{equation}

We should that at those frequencies, the strength of the magnetic field could be constraint using the observed flux, which corresponds to
\begin{multline}
    B'_i \lesssim 125 {\, \rm G} \, \bigg(\frac{\mathcal{D}_i}{3}\bigg)^{-7/3}  \bigg(\frac{\nu_{\rm syn}^{\rm ob}}{7 \, \rm GHz}\bigg)^{-1/3}  \bigg(\frac{\nu L_{\nu, \rm syn }^{\rm ob}}{10^{41} \, \rm erg \, s^{-1}}\bigg)^{2/3}  \\
    \bigg(\frac{K_e'}{10^{7}\rm \, cm^{-3} }\bigg)^{-2/3}\,.
\end{multline}
Our results show that the main contribution from the inner blob is at radio wavelength and VHEs via synchrotron emission by primary electrons and photohadronic processes. 

\paragraph{photopion pairs:} We estimate the synchrotron emission using Eqs.  (\ref{eq_plasma_energy_redshift}) and (\ref{eq_Esyn_photopion}). In this case, the synchrotron energy would be above
\begin{equation}
    E_{\rm syn,p\pi} \gtrsim 36 \, {\rm keV} \,  \Gamma_{\rm rel}^{2} \, \left(\frac{B_i'}{100 \rm G}\right) \, \left(\frac{\mathcal{D}_i}{3}\right)  \, .
\end{equation}

The full parameters used for describing the quiescent state are listed in Tables \ref{tab_model_outer} and \ref{tab_model_inner} for the outer and inner blob, respectively. The resulting spectrum is shown in Figure \ref{fig_SED_quiescent}. 
\subsection{Flaring activity}

Mrk 501 exhibited two flares in May 2009. The first flare started on 2009 May 1 (MJD 54952) and was detected by \textit{Whipple} 10m telescope for 2.3 hours above $>300 \rm \, GeV$. During the first 0.5 hours (i.e, at MJD 54952.37) the flare reached its maximum flux $\sim 4.5 \rm \, C.U.$. After that, VERITAS started the observation at MJD 54952.41 (1.4 hours after the peak), and the flux was decreased to 1.5 C.U. There was no significant increase in X-ray flux which was pointed out as a tentative VHE orphan flare \citep{Ahnen2017A&A...603A..31A}. However, a hardening spectrum was reported by Swift/XRT. They also pointed out that the hardening must have been due to a shift of the synchrotron bump towards higher energies. Furthermore, an interesting feature is the optical polarization degree increases during this flare.  

The MAGIC telescope observed the second flare; on May 22 (MJD 54973), it was observed and reported a flux increased by 3 times the low flux level.

On May 24 (MJD 54975), observations during $\sim 3 \rm \, hour$ measured a decreased flux to $\sim$ 0.5 C.U. Concerning the X-ray band, this flare shows an increase in a factor of $\sim 2$ in the band of 2-10 keV respect to the average value.  Against the first flare, the spectrum did not present a hardening or another change in the X-ray spectral shape.

We consider the same values of variability timescales for two flares as suggested in  \cite{Ahnen2017A&A...603A..31A}. They approximated as $t_{\rm var}^1 \sim 0.5 \rm \, hour $ and $t_{\rm var}^2 \sim \rm \, few \, days $ for the first and second flares, respectively.
The result reported by \citep{Ahnen2017A&A...603A..31A} showed that the spectral indexes of the VHE spectrum during both flares were harder than the ones observed during the quiescent state. In our model, that emission is produced in the inner blob, and we cannot only assume that $L_{\rm keV}$ or $\Gamma_i$ take higher values because, as we discussed in section \ref{sec_flare_state_model}, these only increase the flux without changing the spectral shape.  Therefore, we need to assume that $L_{\rm keV}$ or $\Gamma_i$ increase, and simultaneously the spectrum becomes harder due to a reacceleration process. Here, we do not consider the case when $L_{\rm keV}$ increases because the process is more related to the properties of the accretion disc instead of the blob, and we are only interested in explaining the flares using the changes in the properties of the blobs.

We consider two scenarios to produce the VHE flares by the inner blob, which will be used to model the two flares of Mrk 501 in May 2009.

\paragraph*{a. MODEL A.} The inner blob accelerates, increasing its Lorentz factor, $\Gamma_i$. In contrast, simultaneously, the electron/proton distribution in the blob becomes harder, conserving the particle number $N_{\rm flare}=N_{\rm quiescent}$.
\paragraph*{b. MODEL B.} A new electron/proton distribution is injected in the inner blob, which increase the particle number $N_{\rm flare}>N_{\rm quiescent}$.

In both models, as we did for a quiescent state, we assume that electrons are injected in the inner blob with $\gamma_{e, \rm min}' = 1$ and protons with $\varepsilon'_{p, \rm min}= 1 \rm \, GeV$. The values of the magnetic field determine the value of $\gamma_{e,\rm br}'$ and $\varepsilon_{p, \rm max}'$ given by Eqs. (\ref{eq_gamma_br}) and (\ref{eq_gamma_max}), respectively.   Furthermore, electrons and protons are injected with the same spectral indexes. Moreover, for electrons, we assume $\alpha_{e,2}= \alpha_{e,1}+1$.

Finally, the outer blob could also suffer some changes triggering observable signatures in the X-ray band as presented during both flares. Because two flares present different X-ray behavior, we apply different assumptions for each flare's outer blob. The choice will be explained in the following.

\subsubsection{First flare (MJD 54952)}

The first flare is tentatively identified as an orphan flare. Since the variability timescales are of the order an hour, using the causality argument $R'_1 \lesssim 5 \times 10^{13} \, {\rm cm} \, \mathcal{D}_i \, (t_{\rm var}/0.5 \, \rm hr)$, the emission zone must be a very compact one. Therefore, we adopt the inner blob's size $R'_1 = 1 \times 10^{14} \, {\rm cm}$ based on variability observations here.
We assume VHE gamma-ray flares are produced in the inner blob. Therefore, we apply MODEL A and MODEL B using the parameters of Table \ref{tab_model_inner}, and the resulting spectrum is plotted in Figure \ref{fig_SED_result}.
%
%

Furthermore, this flare presents a hardening in the X-ray spectrum, which could be attributed to a shift of the synchrotron peak produced by the outer blob.   This shift could be produced by an increase of $\gamma_{e, \rm br}''$ in the outer blob resulting as a consequence of the decrease in the magnetic field (see Eq. \ref{eq_gammae_br_outer}). On the other hand, we note that in Figure \ref{fig_SED_quiescent}, for quiescent state, the synchrotron emission of photopion secondary pairs peaks around $\nu_{\rm syn, p\pi}\sim 10^{19} \, \rm Hz$.  This contribution increases proportional to VHE gamma-rays flux, as we mentioned in section \ref{sec:MODEL_quiescent}.  In this work, during the first flare, we assume the emission of photopion pairs causes the hardening of the X-ray spectrum.

The result is plotted in Figure \ref{fig_SED_result} and the corresponding parameters are listed in Table \ref{tab_model_outer}.
Our result shows a good description of the flare; models A and B offer similar descriptions, and the main differences correspond to the parameters used.

\subsubsection{Second Flare (MJD 54973)}
The second flare differs from the first flare, mainly due to the variability timescale; the second flare is longer than the first one.  This would indicate that the emitting region of the second flare could be greater than the first one; which using the causality argument constraint the emitting region size by $R_2'\lesssim 5 \times 10^{15} \, {\rm cm} \, \mathcal{D}_i \, (t_{\rm var}/2 \, \rm day)$. Nevertheless, another option is that the process that triggers the second flare takes more time than the first flare. Assuming this alternative, we assume the size of the emitting regions is similar, then $R_2' \approx 1 \times 10^{14} \, \rm cm$.
%
The result is shown in Figure \ref{fig_SED_result}b. Our result shows that MODEL A and B cannot successfully explain the the X-ray and the VHE gamma-ray flux spectrum. This could suggest that the X-ray flux comes from another region different from the inner blob. Therefore, we consider a third scenario named ``MODEL C".


\paragraph*{c. MODEL C.} The increase in X-ray flux is the product of the outer blob's acceleration, and the VHE gamma-ray flux results from particle injection into the inner blob. 

The result of this case is plotted in Figure \ref{fig_SED_result}b. MODEL C offers a better solution than MODEL A and MODEL B for the second flare
but with the cost of accelerating protons at EeV energies. (see Table \ref{tab_model_inner}).

\section{Discussion and Conclusion}

This work presents a two-zone model to explain the quiescent and two flare activities during May 2009 in Mrk 501. We showed that the quiescent state could be well described by the recent model proposed by AR2022 with similar parameter values as it found for the six best-known hard-TeV BL Lacs. In contrast with AR202, we consider the emission of secondary pairs produced via the photopion process. This emission peaks in the hard X-ray band and could produce an observable signature during intense VHE gamma-ray flares. Our result indicates that the model offers an excellent solution to explain the SED of Mrk501 during a quiescent state.

Furthermore, we implemented the AR2022 model to explain the flare activities presented on Mrk 501. We considered the episodes presented in 2009, May 1st and 22nd. We discussed many possibilities that could cause a flaring episode which could be triggered mainly by three possibilities: i) an increase of the pair-plasma luminosity, ii) changes in the electron/proton distribution in order to produce a large $L_{e,p}$, and iii) the acceleration of the emission zone. The first possibility is that only the efficiency is enhanced and does not change the resulting spectral shape. Meanwhile, the second and third ones produce changes in the shape of the resulting spectrum, producing a hardening of the spectral index or shifting the peak energy, respectively. {\bf We estimate that the kinetic luminosity is to be $ L_{\rm k}\simeq (1.6-42) \times 10^{45} \, \rm erg \, s^{-1}$. We note that the kinetic luminosity as expected is well above that lower limit of radio galaxies (Fanaroff-Riley I) \citep{Punsly2011_ApJ...735L...3P}}. Although the two flares presented some significant differences, they could be successfully fitted with our model. The VHE emission of both flares, first and second, are well explained by the same mechanism. VHE gamma-rays are produced by interacting photons of the pair plasma and protons inside the inner blob. Therefore, we assumed an increase in proton luminosity that triggers the flare. 
The flare produced in the inner blob presents another unique signature in the keV band, which could be tested with observations. We applied two scenarios (MODEL A and B) that produced a similar fit. The first assumes that an acceleration of the inner blob and a re-acceleration of electrons and protons triggers VHE gamma-rays. Meanwhile, the second model assumes electrons and protons are injected with new distribution in the inner blob.   Both cases required similar proton/electron luminosity and magnetic field, but a small value of equipartition parameter for the second model.  A third scenario (MODEL C) was applied for the second flare, which offers the best fit instead of MODEL A and B. This case involves activities in the outer blob and the acceleration of protons in the inner blob in the EeV regime. While the first flare is well-fitted by MODEL A, the second one is favored by MODEL C.

Both flares also showed hints of X-ray activities, where the emission of only either the outer or inner blob cannot wholly explain the observed spectrum. In the first flare, only a hardening in its X-ray spectrum occurs. The emission of photopion pairs interprets this hardening, and there is no need to invoke activity in the outer blob. On the other hand, the second flare increased the X-ray flux without changing the spectrum. Against first flares, this flux cannot be interpreted by secondary pairs. Therefore, we interpreted this as the acceleration of the outer blob, and the results preserve the quiescent state's equipartition value, $U_B/U_e \approx 0.13$.

A significant result for our model is that the equipartition parameter decreased in the inner blob during both flares. This could suggest that flares are triggered by a transfer of magnetic energy to the kinetic energy of particles.
Our model explained two VHE flares episodes, assuming a harder electron/proton distribution inside the inner blob. In addition, our result did not show a significant difference between the injection of a new harder electron/proton distribution or the reacceleration of the same population.

An exciting feature observed during the first flare is the increase in the optical polarization degree, which our model cannot explain. Possibly, the synchrotron emission of primary electrons in the inner blob shifts to the optical band, which contributes to the flux adding up more polarized photons. That polarization flare was not observed during the second flare, suggesting that the nature of both flares was distinct. 

It is essential to mention that, in our model, the produced neutrino flux peaks around $\gtrsim$ 1 TeV \citep{AguilarRuiz_2023JHEAp..38....1A}. However, our predicted spectral shape differs from the IceCube Collaboration's, especially for the blazar TXS 0506 +056, which spectrum is well described by an unbroken power-law with a spectral index of $\approx 2.2$ and does not necessary peaking at TeV energies \cite{IceCube2018Sci...361..147I,IceCube2020PhRvL.124e1103A}. Our scenario agrees better with the case of NGC 1068,  where the spectrum is a softer one $\approx 3.2$, and the most contribution comes from the energy range of 1.5-15 TeV, and out of that range data cannot strongly constraint the properties of the inferred flux \cite{IceCube2020PhRvL.124e1103A,IceCube_2022Sci...378..538I} . Nevertheless, another option might be one where the total spectrum is composed of two neutrino components, one at TeV energies and another at higher energies.

Furthermore, multiwavelength observations with simultaneous neutrino observations during future intense flaring activities in blazars may provide additional constraints to our model.  Recently, \citep{Kun_2021ApJ...911L..18K,Halzen2022arXiv220200694H} pointed out that high-energy neutrinos are produced when the gamma-ray fluxes are suppressed due to increased seed photons, which enhance photopion efficiency and pair creation process. In our scenario, seed photons are the 511 keV annihilation line, which strongly attenuates the gamma-ray flux from MeV to GeV energies.

Our model produced an essential signature near 1 MeV, which could be tested with the observation by future telescopes such as AMEGO or eASTROGAM during flare episodes in BL Lacs. Further, multiwavelength campaigns that can be observed simultaneously during flare episodes are the key to testing the model.

\section{Acknowledgements}

We appreciate the referee's insightful remarks and suggestions for enhancing the quality of this work. We are also grateful to Peter Veres and Antonio Marinelli for useful discussions.  This work is supported  by UNAM-DGAPA-PAPIIT  through  grant  IN106521.

\bibliography{apssamp}

\clearpage

\begin{table}
\caption{Parameters used to model the states of Mrk 501 on 2009 may with the outer blob.}
{
\begin{tabular}{lccccc }
\hline 
& \multicolumn{4}{c}{\textbf{ Outer blob }}
\\
\hline
\hline
& \multicolumn{1}{c}{\textbf{ Quiescent }}     
& \multicolumn{1}{c}{\textbf{ Flare 1}}         
& \multicolumn{1}{c}{\textbf{ Flare 2}}
&
\\
& \multicolumn{1}{c}{\textbf{ }}     
& \multicolumn{1}{c}{\textbf{ }}         
& \multicolumn{1}{c}{\textbf{ (C) }}
\\
\hline 
 $\Gamma_o \,$ 
 & $5$ & $5$ & $5.5$
 \\
 $\mathcal{D}_o$ 
 & $10$ & $10$ & $11$
 \\
 $R'' \rm \, [10^{16} \, cm]$   
 & $1.2 $ & $1.2 $ & $1.3 $
 \\
 $B'' \, \rm [G]$   
 & $0.23$ & $0.23$ & $0.23$ 
 \\
$K^{\prime\prime}_e \, {\rm [10^{4} \, cm^{-3}] }$ 
& $2.2 $ & $2.2 $ & $2.2 $
\\
${\gamma''}_{e,\rm min} $ 
&$70$ & $70$ & $70$
\\
$\gamma''_{e,\rm br} [10^5] $ 
&$1 $ & $1 $ & $1$
\\
$\gamma''_{e,\rm max} [10^6]$ 
& $ 5 $ & $5 $ & $5 $
\\
$\alpha_{e,1}$ 
& $2.3$ & $2.3$ & $2.3$
\\
$\alpha_{e,2}$ 
& $3.2$ & $3.2$ & $3.2$
\\
$L_e^{\rm ob} \; [ \rm 10^{44} \, erg \, s^{-1} ]$ 
& $1.3 $ & $1.3 $ & $1.8 $
\\
$L_B^{\rm ob} \; [ \rm 10^{43} \, erg \, s^{-1} ]$ 
& $1.8$ & $1.8$ & $2.4$
\\
$U_B/(U_e+U_p)$ 
& 0.13 & 0.13 & 0.13
\\
\hline
\end{tabular}}\label{tab_model_outer}
\end{table}

\begin{table}
\caption{Parameters used to model the flaring states in May 2009 in Mrk 501 with the inner blob.}
{
\begin{tabular}{l cc cc cc cc }
\hline 
& \multicolumn{6}{c}{\textbf{ Inner blob }}
\\
\hline
\hline
   & \multicolumn{1}{c}{\textbf{ Quiescent }}
   & \multicolumn{2}{c}{\textbf{ Flare 1}} &
   & \multicolumn{3}{c}{\textbf{ Flare 2}}  
\\
   & \multicolumn{1}{c}{\textbf{}} 
   & \multicolumn{1}{c}{(A)}      
   & \multicolumn{1}{c}{(B)} &      
   & \multicolumn{1}{c}{(A)}
   & \multicolumn{1}{c}{(B)}
   & \multicolumn{1}{c}{(C)}
\\
\hline 
 $\Gamma_i$ 
 & $1.3 $ & $2.0$ & $1.5 $ && $2.0$ & $1.3 $ & $1.3$
 \\
 $\mathcal{D}_i$ 
 & $2.1$ & $3.7$ & $2.9$ && $3.7$ & $2.1$ & $2.1$
 \\
 $R'_i \rm \, [10^{14} \, cm]$   
 & $1$ & $1$ & $1$ && $1 $ & $1$ & $1$
 
 \\
 $B'_i \, \rm [G]$   
 & $100$ & $100$ & $120$ && $80$ & $100$ & $100$ 
 \\
$K'_p \, {\rm [10^{7} \, cm^{-3} \, GeV^{-1}] }$ 
& $4.8$ & $2.1$ & $3.6$ && $2.7$ & $12$ & $4.7$
\\
${\varepsilon'}_{p, \rm min} \; [\rm GeV] $ 
& $ 1 $ & $ 1$ & $ 1$ && $ 1$ & $ 1 $ & $ 1 $
\\
${\varepsilon'}_{p, \rm max} \; [\rm PeV]$ 
& $ 100 $ & $100 $ & $100 $ && $ 100 $ & $100$ & $6 \times 10^3 $
\\
$\alpha_p$ 
& $2.1$ & $1.9$ & $1.8$ && $2$ & $2$ & $1.8$
\\
$K'_e \, {\rm [10^{7} \, cm^{-3}] }$ 
& $4.3$ & $2.0$ & $3.7$ && $2.0$ & $10.0$ & $4.6$
\\
${\gamma'}_{e,\rm min} $ & 
$1$ & $1$ & $1$ && $1$  & $1$ & $1$
\\
$\gamma'_{e,\rm br} $ &
$10$ & $10$ & $10$ && $7$ & $10$ & $10$
\\
${\gamma'}_{e,\rm max} \; [10^5]$ & 
$ 1$ & $1$ & $1$ && $1$ & $1$ & $1$
\\
$\alpha_{e,1}$ 
& 
$2.1$ & $1.9$ & $1.8$ && $2$ & $2$ & $1.9$
\\
$\alpha_{e,2}$ 
& 
$3.1$ & $2.9$ & $2.8$ && $3$ & $3$ & $2.9$
\\
$n'_p=n'_e \, {\rm [ 10^{7} \, cm^{-3}] }$ 
& $2.0 $ & $2.0$ & $4.2$ && $2.0$ & $11.0$ & $4.9$
\\
$L_p^{\rm ob} \; [ \rm 10^{45} \, erg \, s^{-1} ]$ 
& 
$1.6$ & $22.0$ &  $42.0$ && $9.0$ & $8.7  $& $15.0$
\\
$L_e^{\rm ob} \; [ \rm 10^{41} \, erg \, s^{-1} ]$ 
& 
$2.4$ & $8.6$ & $4.8$ && $8.1$ & $7.2$& $3.4$
\\
$L_B^{\rm ob} \; [ \rm 10^{42} \, erg \, s^{-1} ]$ 
& 
$1.0$ & $6.0$ & $2.7$ && $3.8$ & $1.0$& $1.0$
\\
$U_B/(U_e+U_p) [10^{-4}]$ 
& 
$0.7$ & $3.0$ & $0.7$ && $4.0$ & $1.0$ & $0.7$
\\
\hline
\end{tabular}}\label{tab_model_inner}
\end{table}

\begin{figure}
\centering
\includegraphics[width=\linewidth]{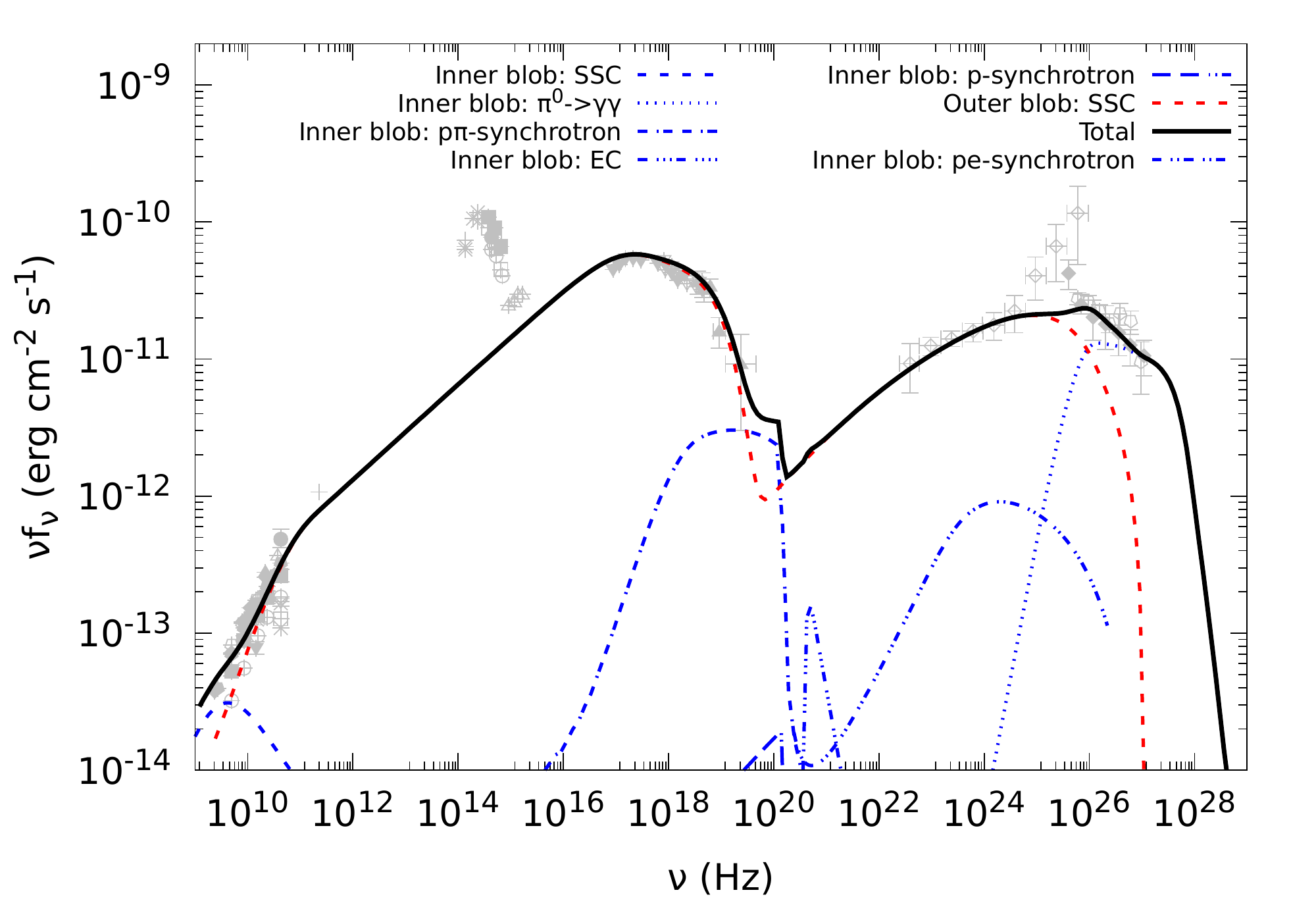} 
\caption{The quiescent state is modelled using a two-zone leptohadronic model. We show the outer blob emission via SSC and the inner blob emission via SSC, proton-synchrotron, photopair and photopion processes. The data for the quiescent state is shown in grey points while the solid black line shows the total flux obtained by our model.}%
  \label{fig_SED_quiescent}
\end{figure}

\begin{figure*}
\begin{minipage}[b]{0.5\linewidth}
\centering
\includegraphics[width=\linewidth, height=7cm]{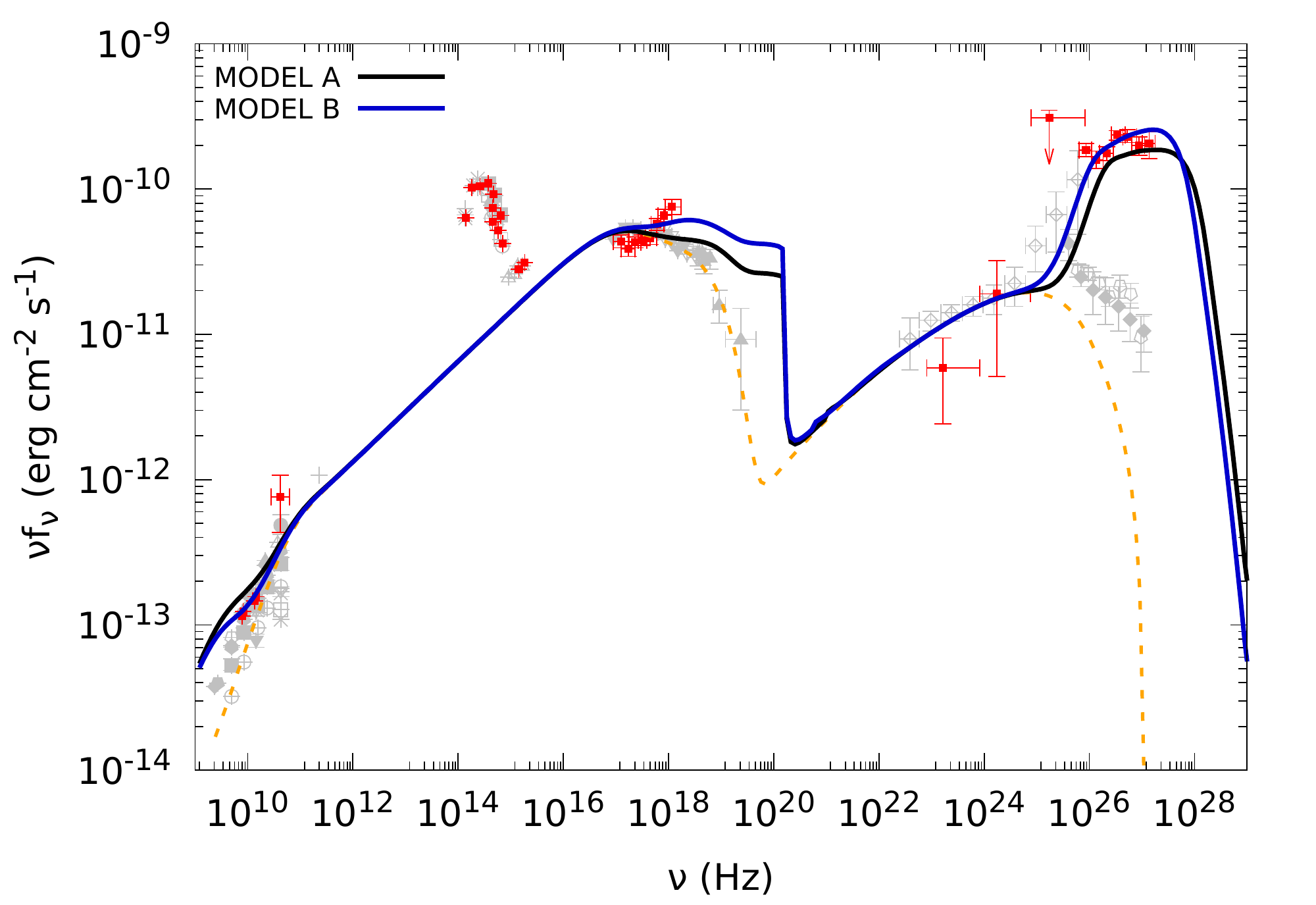}
\end{minipage}\hfill 
\begin{minipage}[b]{0.50\linewidth}
\centering
\includegraphics[width=\linewidth, height=7cm]{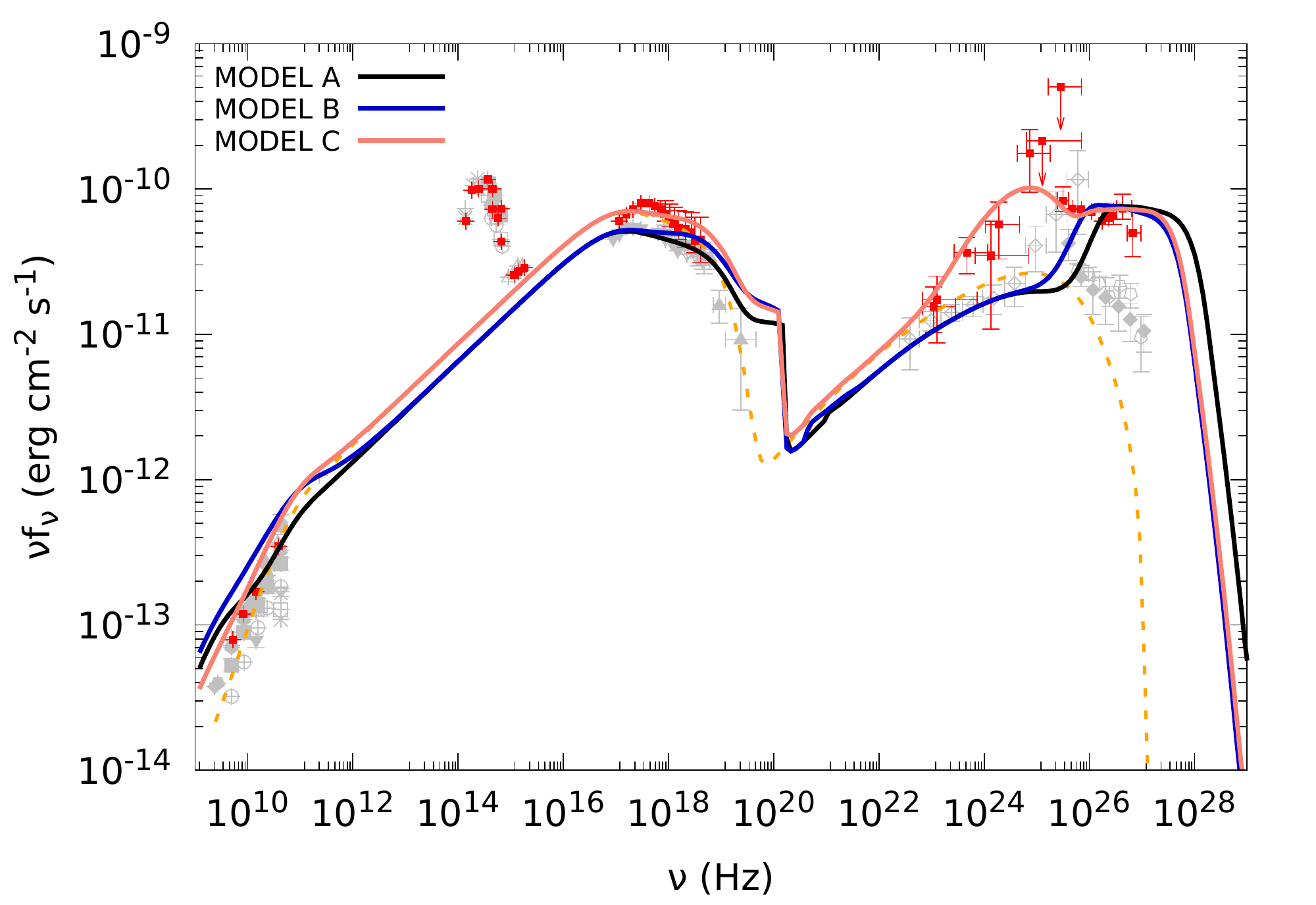} 
\end{minipage}
\caption{The result of two flaring activities using our model. 
The SSC flux in the outer blob is shown with orange dashed lines
The quiescent state is shown in grey points, while red points represent the flaring states. \textbf{a)} The first flare on 2009 May 1 (MJD 54952) \textbf{b)} The same as the (a), but applied to the second flare on May 24, 2009 (MJD 54973).}%
  \label{fig_SED_result}
\end{figure*}
\end{document}